\documentclass[11pt]{article}
\pdfoutput=1
\usepackage{jheppub}
\usepackage{tabu}
\usepackage[vcentermath]{youngtab}
\usepackage[usenames,dvipsnames,table]{xcolor}
\usepackage{graphicx,amsmath,amssymb,amsthm,multirow,array,bm,bbm,esint}
\usepackage[mathscr]{eucal}
\usepackage[bbgreekl]{mathbbol}
\usepackage{epsf,amsfonts}
\usepackage{slashed}
\usepackage[numbers,sort&compress]{natbib}
\usepackage{minibox}
\usepackage{rotating}
\usepackage{pdflscape}
\usepackage{array,tikz-cd}
\usepackage{subfigure}
\usepackage{pgfplots}
\pgfplotsset{compat=1.15}
\usepackage{mathrsfs}
\usetikzlibrary{arrows}
\usetikzlibrary{backgrounds}
\usepackage{slashed}

\usepackage{import}
\usepackage{pdfpages}
\usepackage{transparent}

\definecolor{rust}{rgb}{0.8,0.2,0.2}

\definecolor{red}{rgb}{0.9,0.1,0.1}

\def\VH#1{{\color{cyan}{#1}}}

\definecolor{rust}{rgb}{0.6,0.2,0.2}

\definecolor{labelkey}{rgb}{0.4,0.4,0.4}

\title{Theory dependence of black hole interior reconstruction and the extended strong subadditivity}

\author{Sitender Pratap Kashyap,}
\author{Roji Pius,}
\author{Manish Ramchander}

\affiliation[a]{
The Institute of Mathematical Sciences, IV Cross Road, C.I.T. Campus, Taramani, Chennai, India 600113}

\affiliation[b]{
Homi Bhabha National Institute, Training School Complex, Anushakti Nagar, Mumbai, India 400094}

\emailAdd{sitenderpk@imsc.res.in}
\emailAdd{rojipius@imsc.res.in}
\emailAdd{manishd@imsc.res.in}

\abstract{ 
An AdS eternal black hole in equilibrium with a finite temperature bath presents a Hawking-like information paradox due to a continuous  exchange of radiation with the bath. The non-perturbative gravitational effect, the replica wormhole,  cures this paradox by introducing a non-trivial entanglement wedge for the bath after Page time. In this paper, we analyse the theory dependence of this non-perturbative effect by randomising the boundary conditions of some of the bulk matter fields. We explicitly analyse this in  JT gravity by introducing a  matter CFT in the AdS region with random boundary conditions at the AdS boundary that are drawn from a distribution. Using the island formula and the extended strong subadditivity due to Carlen and Lieb, we show that at late times the black hole interior is contained inside the entanglement wedge of a reference Hilbert space that encodes the information about the random boundary conditions. Consequently, the reconstruction of the black hole interior from  the radiation, in particular the region near the singularity, requires a detailed knowledge of the theory. 
}

\begin{document}
\maketitle

 
\section{Introduction}
\label{sec:intro}

The resolution of Hawking information paradox \cite{Hawking:1976bpgc} of an evaporating black hole in AdS requires introducing a non-trivial entanglement wedge that contains the black hole interior after the Page time for the Hawking radiation \cite{Penington:2019npb, Almheiri:2019qdq,  Almheiri:2019hni, Almheiri:2020cfm}.  This appearance of a non-vanishing entanglement wedge for Hawking radiation after the Page time is result of a non-perturbative gravitational effect, the replica wormhole. Usually, non-perturbative effects are highly theory dependent. The energy spectrum of a black hole \cite{Cotler:2016fpe, Saad:2018bqo},  the S-matrix of a black hole that determines  its formation, and evaporation of the black hole \cite{Polchinski:2015cea} are some of the  examples of such quantities.  Since the ability of late Hawking radiation to reconstruct the black hole interior is a non-perturbative effect,  it is natural to suspect that the interior reconstruction might depend on the details of the theory.  \par

An AdS eternal black hole in equilibrium with a finite temperature bath also comes with an information paradox very much like the Hawking information paradox. The paradox, the unbound growth of bath entropy, is due to the continuous exchange of the Hawking radiation and the radiation from the bath. The resolution of this paradox also requires introducing a non-trivial entanglement wedge for the bath radiation after the Page time \cite{Almheiri:2019hni}. Compared to an evaporating black hole, a black hole in equilibrium with a finite temperature bath is a more convenient setup for studying the theory dependence of the reconstruction from radiation due to the absence of any backreaction. For an eternal AdS black hole with no matter escaping the AdS boundary, it is already  demonstrated in \cite{Almheiri:2021jwq} that the bulk reconstruction of the interior of the black hole is highly theory dependent at late times. They achieved this by making the boundary conditions of the bulk matter fields random, and  showing that the reference Hilbert space that encodes the information about this randomness possesses a non-trivial entanglement wedge that contains the black hole interior including the region near the singularity.\par

 In this paper, by following \cite{Almheiri:2021jwq}, we analyse the dependence of the interior reconstruction using the bath radiation on the boundary conditions of the bulk matter fields. For this we consider a JT gravity black hole in equilibrium with a non-gravitating bath at finite temperature and introduce a matter CFT coupled to gravity in the black hole region having reflecting boundary conditions for the fields in it. We assume that that these boundary conditions are drawn from a probability distribution.  We denote the probability for the $i^{th}$ field to have a boundary condition $J_i$  as $P(J_i)$\footnote{In the SYK picture,  $\mathbf{J}=\{J_1,\cdots, J_i,\cdots\}$  would correspond to the random couplings that appear in the SYK Hamiltonian.}.   Then the black hole density matrix defined using the Euclidean path integral will depend on the boundary conditions $\mathbf{J}=\{J_1,\cdots, J_i,\cdots\}$  of the bulk CFT matter fields and the associated probability distribution  $P(\bold{J})=\{P(J_1),\cdots, P(J_i),\cdots\}$.  The purification of this  density matrix   requires the bath Hilbert space $\mathcal{H}_{bath}$ and also the introduction of an environment, an auxiliary reference Hilbert space $\mathcal{H}_{journal}$, which is referred as  the `journal' Hilbert space.  The journal Hilbert space encodes the information about the boundary conditions of the bulk CFT matter fields. Therefore  the dependence of the black hole interior reconstruction  using the bath radiation on the  the boundary conditions of the matter fields can be characterised  by determining the entanglement wedge of the journal \cite{Dong:2016eik}. The goal of this paper is the  determination of the entanglement wedge of the journal for this setup. 
\par

The physical significance of this problem was already discussed in \cite{Almheiri:2021jwq} for an evaporating black hole which has two additional systems other than the black hole, the Hawking radiation and the journal. Our setup similarly has two additional systems other than the black hole, the bath radiation and the journal. However, the absence of  backreaction makes our setup more convenient for analysing the same problem.  After the black hole Page time the bath radiation and the journal strive for the ownership of the interior of the black hole. The winner is expected to be decided by the rate of the entropy growth of the two systems. As it is computed in this paper, initially the entropy of bath radiation grows linearly and the journal entropy grows logarithmically. Therefore, at first the bath radiation is expected to capture the black hole interior. The non-triviality is in figuring out whether the ownership of the interior is ever transferred to the journal. If the bath radiation retains the interior forever, then it means that the interior reconstruction is insensitive to details of the bulk theory, which suggests that the unknown details of the bulk theory can be determined by making measurements on the bath radiation.  If the ownership is transferred to the journal at a late time, then it implies that interior reconstruction is theory dependent. The main result of this paper is that the bath radiation transfers ownership of the black hole interior to the journal at a later time.  \par 

Let us briefly delineate how we proceed.  We determine the entanglement wedge of the subsystems, black hole, bath and journal, by demanding that the entropies of these  subsystems  satisfy  all the constraints imposed by unitarity. There are two such constraints, first is that the von Neumann entropy of any subsystem must be less than its thermal entropy and the second is that the entropies of the subsystems must satisfy the extended strong subadditivity (eSSA) due to Carlen and Lieb \cite{carlen2012bounds}.  The first constraint demands that the von Neumann entropy $S(\rho_{BH})$ of the eternal black hole density matrix $\rho_{BH}$ must be less than $2S^{0}_{BH}$, where  $S^{0}_{BH}$ is the Bekenstein-Hawking entropy of the one side of the eternal black hole. Since black hole has only a finite number of degrees of freedom,   $S^{0}_{BH}$ is  finite. This demands that the entanglement entropy of the black hole must not be an ever-growing function of boundary time.  However, the black hole entropy obtained by the replica computation without including any Euclidean wormhole contribution becomes more than $2S^{0}_{BH}$  after the black hole Page time. This violation of thermal entropy bound can be cured by removing the black hole interior from the combined entanglement wedge of the boundary CFTs dual to the eternal black hole. The resulting generalised entropy of the black hole after the Page time saturates the thermal entropy bound $2S^0_{BH}$.  On the contrary, the entropy $S(\rho_{bath})$ of the bath density matrix $\rho_{bath}$ and the entropy  $S(\rho_{journal})$ of the journal density matrix $\rho_{journal}$ can have unbounded growth due to the infinite number of degrees of freedom they possess. Hence one naively expects that the interior of the black hole after the Page time may be co-owned by the bath and the journal.\footnote{By co-ownership, we mean that neither the bath nor the journal individually owns the interior, only the combined system does.}  However, using the second constraint we argue below that this is not true. \par

  If the state of the combined system of the black hole, the bath and the journal  is pure, then the eSSA after the black hole Page time states that 
\begin{equation}
\label{improvedSSABHJB11}
S(\rho_{bath})+S(\rho_{journal})-2S^0_{BH}\geq 2~max\left\{S(\rho_{journal})-2S^0_{BH}, S(\rho_{bath})-2S^0_{BH},0\right\}.
\end{equation}
Using the replica trick we can compute $S(\rho_{bath})$ and $S(\rho_{journal})$. In the absence of any non-trivial islands, $S(\rho_{bath})$ and $S(\rho_{journal})$ grow linearly and logarithmically respectively with respect to the boundary time.  Due to the larger growth of bath entropy, the eSSA takes the following form right after the black hole Page time
\begin{equation}
\label{improvedSSABHJB12}
S(\rho_{bath})-S(\rho_{journal})\leq 2S^0_{BH}.
\end{equation}
This inequality can be satisfied only if the bath owns an island that contains the black hole interior after black hole Page time.  After including such an island, $S(\rho_{bath})$ becomes $2S^0_{BH}$.  However, at a later time the logarithmic growth of $S(\rho_{journal})$ makes it larger than $2S^0_{BH}$. At this stage the eSSA takes the following form
\begin{equation}
\label{improvedSSABHJB13}
S(\rho_{journal})-S(\rho_{bath})\leq 2S^0_{BH}.
\end{equation}
Clearly, this inequality is satisfied until  $S(\rho_{journal})$ becomes $4S^0_{BH}$.  Subsequently, in order to satisfy the eSSA, the bath must transfer the ownership of the black hole interior to the journal. Introducing such a non-trivial entanglement wedge that contains the interior of the black hole makes the rate of the entropy growth of the journal same as that of the bath and saturates the eSSA, thus restoring unitarity.  This implies that the reconstruction of the black hole interior using the bath radiation at late times requires the complete description  of the theory which includes  specifying the boundary conditions of all the fields at the AdS boundary. \par

The paper is organised as follows. In section \ref{sec:setup}, we briefly describe the setup which is an eternal AdS$_2$ black hole in equilibrium with a finite temperature non-gravitating bath with two kinds of matter, one having transparent boundary conditions along the boundary of the gravitational region, and another having random reflecting boundary conditions drawn from a distribution. In section \ref{sec:entwedge of journal}, we determine the entanglement wedge of the random boundary conditions to characterise the theory dependence of the black hole interior reconstruction using the bath radiation. In section \ref{conclusion}, we reiterate the result obtained and briefly touch upon some of the future directions that deserve immediate attention.

\section{The setup}
\label{sec:setup}

Consider a  black hole solution of JT gravity with inverse temperature $\beta$ coupled to a bath having the same temperature. We assume that gravity is absent in the bath. We introduce two CFTs into this spacetime. They will be referred as  CFT$_1$ and CFT$_2$. The CFT$_1$ has  central charge $c_1$ and  CFT$_2$ has central charge $c_2$. The CFT$_2$ is restricted to the gravitating AdS$_2$ region, while the CFT$_1$ lives in the full spacetime, which is the AdS$_2$ and the bath region together. This is done by setting  transparent boundary conditions for fields in CFT$_1$ and reflecting boundary conditions for fields in CFT$_2$.   Both the CFTs are coupled to the metric in the gravitational region. However, they are not coupled to the dilaton field. This makes the black hole spacetime locally AdS$_2$, even though gravity is dynamical in the black hole region. We also  assume that CFT$_1$ and CFT$_2$ do not directly interact with each other.    An additional feature of CFT$_2$  is that the boundary conditions of the fields in this theory are drawn from a distribution. In the dual holographic side this arises from ``unknown couplings" whose information is present in the aforementioned system called journal \cite{Almheiri:2021jwq, Qi:2021oni, Renner:2021qbe}.  \par

\subsection{Black hole in equilibrium with a bath}
The black hole solutions of JT gravity have been used widely \cite{Almheiri:2019psf, Almheiri:2019yqk, Almheiri:2021jwq, Almheiri:2019qdq, Goto:2020rwe} as a toy model for studying black hole evaporation. For detailed reviews see \cite{Sarosi:2017syk, Dmitrii:2020ped, Engelsoy:2016ibh, Almheiri:2014mbh}.  The action for JT gravity coupled to  CFT$_1$ is given by
\begin{align}
	\label{JTaction}
    S=&\frac{\phi_0}{16\pi G_N}\left[ \int_{\cal{M}}d^2x\sqrt{-g}\, R+2\int_{\partial \cal{M}}K\right]\\
    &+\frac{1}{16\pi G_N}\left[ \int_{\cal{M}}d^2x\sqrt{-g}\phi\left(R+\frac{2}{\ell^2_{AdS}}\right)+2\int_{\partial \cal{M}}\phi_bK\right]+I_{\text{CFT}_1}[g]\nonumber
\end{align}
where $R$ is the Ricci scalar of the spacetime $\mathcal{M}$, $K$ is the trace of the extrinsic curvature of the boundary  $\partial \mathcal{M}$, $\phi_b$ is the boundary value of the dilaton field and  $\ell_{AdS}$ is the AdS radius which we will set to 1. Also, $G_{N}$ is the Newton's constant and $\phi_0$ is a constant that sets the extremal entropy to be $\frac{\phi_0}{4G_N}$. The specific manner in which $\partial\mathcal{M}$ is carved out of pure AdS$_2$  is responsible for breaking the reparametrization symmetry and it gives rise to non-trivial bulk dynamics \cite{Maldacena:2016upp}. It is important to note that $I _{\text{CFT}_1}$ does not couple to dilaton $\phi$. Consequently the theory satisfies the constraint $R = -2$ and hence the geometry is locally same as that of pure AdS$_2$.

To describe  the boundary, we use the Poincare patch, whose metric is
\begin{equation}
ds^2 = \frac{-dt^2 + dz^2}{z^2} = \frac{- 4dx^+ dx^-}{(x^+ - x^-)^2} 
\label{Poincare_metric}
.\end{equation}
where $x^\pm = t \mp z$. The standard boundary conditions on the cut out are imposed to be
\begin{equation}
\label{boundarycond}
g_{uu}\vert_{bdy}=\frac{1}{\epsilon^2}=\frac{1}{z^2}\left(-\left(\frac{dt}{du}\right)^2+\left(\frac{dz}{du}\right)^2\right){\Biggl|_{bdy}} \qquad \phi \vert_{bdy}=\phi_b=\frac{ \phi_r}{\epsilon}.
\end{equation}
Here $u$ is the boundary time.  In these coordinates, the boundary is given by $t = f(u), z = \epsilon  f'(u)$, where the gluing function $f(u)$ is obtained from the energy balance equation as follows. The ADM mass of the gravitating region is
\begin{eqnarray}
	M(u)=-\frac{\phi_r}{8\pi G_N} \{f(u),u\}.
\end{eqnarray}
Energy conservation requires that the change in above ADM mass equal the net flux of energy across the boundary curve
\begin{eqnarray}
	\frac{dM(u)}{du}=-\frac{d}{du}\left(\frac{\phi_r}{8\pi G_N} \{f(u),u\}\right)=T_{y^+y^+}-T_{y^-y^-}
\end{eqnarray}
which, given the stress tensor profile, can be solved for $f(u)$. Now vary the action with respect to metric in the Poincare patch to obtain the equations for dilaton:
\begin{equation}
\label{dilatoneq}
\begin{split}
2\partial_{x^+}\partial_{x^-}\phi+\frac{4}{(x^+-x^-)^2}\phi&=16\pi G_N\langle T_{x^+x^-}\rangle\noindent,\\
-\frac{1}{(x^+-x^-)^2}\partial_{x^+}\left( (x^+-x^-)^2\partial_{x^+}\phi\right)&=8\pi G_N \langle T_{x^+x^+}\rangle\noindent,\\
-\frac{1}{(x^+-x^-)^2}\partial_{x^-}\left( (x^+-x^-)^2\partial_{x^-}\phi\right)&=8\pi G_N \langle T_{x^-x^-}\rangle\noindent.\\
\end{split}
\end{equation}

Solution to these equations, up to an $SL(2,R)$ transformation, can be written as 
\begin{eqnarray}
	\phi(x^+,x^-)&=&-\frac{2\pi \phi_r}{\beta}\frac{x^++x^-}{x^+-x^-} -\frac{8\pi G_N}{x^+-x^-}\int_0^{x^{-}} dt (x^+-t)(x^- -t)T_{x^- x^-}\nonumber\\
&& +\frac{8\pi G_N}{x^+-x^-}\int_0^{x^{+}} dt (x^+-t)(x^- -t)T_{x^+ x^+}.
\end{eqnarray}
Then the dilaton takes form  \cite{Hollowood2020:ipc, Goto:2020rwe}
	\begin{eqnarray}
		\phi(x^+,x^-)=-\phi_r\left(\frac{2f'(y^+)}{x^+-x^-}-\frac{2f''(y^+)}{f'(y^+)}\right).
        \label{dilaton_gluing_function}
	\end{eqnarray}

\begin{figure}
\centering
\begin{tikzpicture}[scale=.85]
    \draw [black!50!green,dotted, thick](1 mm, 10 pt) (-1,.25)--(-1,4.75) ;
      \draw [black!50!green,dotted, thick](1 mm, 10 pt) (-6,.25)--(-6,4.75) ;
  \draw [color=black!50!violet,thick](1 mm, 10 pt) (-6,4.75)--(-1,.25) node[pos=.7,sloped, above] {$w^+=0$};
 \draw [color=black!50!violet,thick](1 mm, 10 pt) (-6,.25)--(-1,4.75)node[pos=.3,sloped, above] {$w^-=0$};
  \draw [thick,](1 mm, 10 pt) (1.5,2.5)--(-1,.25) node[pos=.5,sloped, below] {$w^-=-\infty$} ;
    \draw [thick,](1 mm, 10 pt) (-6,4.75)--(-3.5,7) ;
    \draw [thick,](1 mm, 10 pt) (-1,4.75)--(-3.5,7) ;
      \draw [thick,](1 mm, 10 pt) (-6,.25)--(-3.5,-1.9) ;
    \draw [thick,](1 mm, 10 pt) (-1,.25)--(-3.5,-1.9) ;
 \draw [thick](1 mm, 10 pt) (1.5,2.5)--(-1,4.75)node[pos=.5,sloped, above] {$w^+=\infty$} ;
   \draw [thick,](1 mm, 10 pt) (-8.5,2.5)--(-6,.25) node[pos=.5,sloped, below] {$w^+=-\infty$} ;
 \draw [thick](1 mm, 10 pt) (-8.5,2.5)--(-6,4.75)node[pos=.5,sloped, above] {$w^-=\infty$} ;
 \draw [color=black!30!blue, very thick]   (-1,.25)  to[out=110,in=-110] (-1,4.75);
 \draw [color=black!30!blue, very thick]   (-6,.25)  to[in=290,out=-290] (-6,4.75);
  \draw [color=black!30!red, dashed,very thick]   (-1,4.75)  to[out=-140,in=-40] (-6,4.75) ;
  \draw  (-3.5,4.25) node[ above] {$w^+w^-=\frac{1}{\theta}$};
 \draw [color=black!30!red, dashed, very thick]   (-1,.25)  to[in=40,out=140] (-6,.25);
   \draw  (-3.5,1) node[ below] {$w^+w^-=\frac{1}{\theta}$};
\end{tikzpicture}
\caption{Eternal black hole in thermal equilibrium with a bath can be described using the $w$-plane.  The right and left Rindler wedge in the $w$-plane describes the right and left side of the black hole coupled to non-gravitating bath having same temperature as that of the black hole. } \label{fig:BHbath1}
\end{figure}
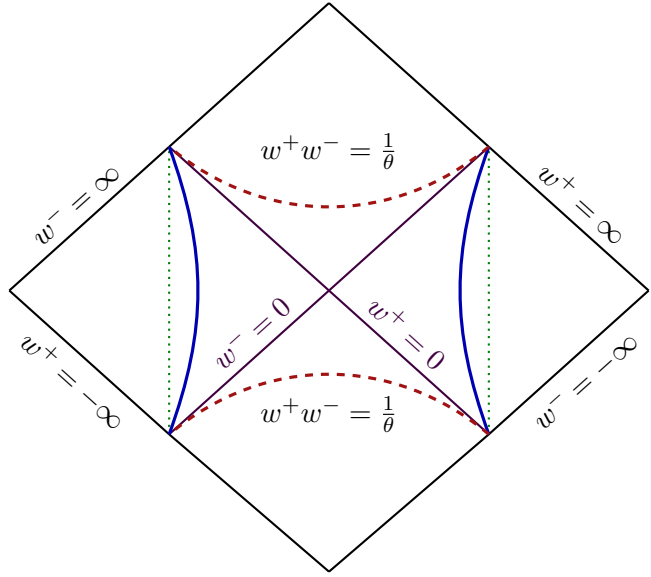

Now we describe how to make use of the above equations to couple the two sides of a black hole solution in the AdS$_2$ to the Minkowski bath having coordinates  $y^{\pm}=f^{-1}(y^\pm)$ and metric
\begin{eqnarray}
	ds^2=-\frac{1}{\epsilon^2}dy^+dy^-.
\end{eqnarray}
 We demand that the black hole is in equilibrium with the bath, there is no net flux and hence 
\begin{eqnarray}
	\partial_u\{f(u),u\}=0.
\end{eqnarray}
A solution that corresponds to a temperature $\frac{1}{\beta}$ is given by 
\begin{eqnarray}
f(u)=e^{\frac{2\pi u}{\beta}}.
\end{eqnarray}
Having solved for the gluing function, we can use it to extend the coordinates $y^{\pm}$ that were earlier defined in the bath region to the gravity region as well via  \begin{equation}
x^\pm = f(y^\pm) = \pm \exp \left( \pm\frac{2 \pi}{\beta } y^\pm \right).
    \label{eq:poincareToRindler}
\end{equation}
Given the map \eqref{eq:poincareToRindler}, the Poincare metric in \eqref{Poincare_metric} becomes 
\[ 
ds^2 =  - \left( \frac{2\pi}{\beta } \right)^2 \frac{dy^+ dy^-}{\sinh^2 \frac{\pi}{\beta }(y^- - y^+)},
\]
and the dilaton profile takes the form 
\[ 
\phi  = \frac{2\pi \phi _r}{\beta } \frac{1}{\tanh \frac{\pi}{\beta } (y^- - y^+)}.
\]
We will however be mostly working in Kruskal-Szekeres coordinates
\begin{align}
\label{kruskal}
w^{\pm}&=\pm e^{\pm\frac{2\pi y_R^{\pm}}{\beta}}=\pm \left(x^{\pm}_R\right)^{\pm 1} \qquad {\text{for~right~side~of~the~glued~geometry}}\nonumber \\
w^{\pm}&=\mp e^{\mp\frac{2\pi y_L^{\pm}}{\beta}}=\mp \left(x^{\pm}_L\right)^{\mp 1} \qquad {\text{for~left~side~of~the~glued~geometry}}.
\end{align}
In these coordinates, the black hole metric takes the form
\begin{equation}
\label{eq:kruskmetric}
ds^2=\frac{4dw^-dw^+}{\left(1+w^-w^+\right)^2},
\end{equation}
  and the dilaton profile becomes
 \begin{equation}
 \label{eq:vacuumdilprofw}
 \phi(w^+,w^-)=\phi_0+\frac{2\pi \phi_r}{\beta}\frac{1-w^+w^-}{1+w^+w^-}. 
 \end{equation}
Therefore, the location of the singularity is given by $w^+w^-=\frac{1}{\theta}$, where $\theta =\frac{2\pi\phi_r-\beta\phi_0}{2\pi\phi_r+\beta\phi_0}$.  Further, the future horizon of the black hole is at $w^-=0$ and past horizon is at $w^+=0$.  Finally, the location of the physical boundary of the black hole geometry that is being glued to the bath is given by $w^+w^-=-e^{\frac{2\pi\epsilon}{\beta}}$. See figure \ref{fig:BHbath1}.

\subsection{Random boundary conditions and the journal}

Let us introduce the second CFT, the CFT$_2$, in the AdS$_2$ region with reflecting boundary conditions for the matter field along its boundary.  Since there is no additional net flow across the interface between bath and the AdS$_2$ region due to CFT$_2$, introduction of CFT$_2$ does not change the geometry of the spacetime. For computational tractability we have chosen a CFT$_2$ that does not interact with CFT$_1$.  Assume that the probability for the $i^{th}$ field in CFT$_2$ to have a boundary condition $J_i$ is  $P(J_i)$.   As mentioned in the introduction, the information of the boundary conditions of CFT$_2$ matter fields $\bold{J}=\{J_1,\cdots,J_i,\cdots\}$ and the associated probability distribution  $P(\bold{J})$ are encoded in the density matrix of the black hole.    This black hole density matrix cannot be purified only by the bath Hilbert space $\mathcal{H}_{bath}$, it also requires introducing an auxiliary reference Hilbert space $\mathcal{H}_{\bold{J}}$ which is referred as  the `journal' Hilbert space.  Let $\{|J_i \rangle_{journal}\}$,  $\{|\psi_k,J_i \rangle_{BH}\}$ and $\{|\gamma_{k'}\rangle_{bath} \}$ be basis for  $\mathcal{H}_{journal}$, $\mathcal{H}_{BH}$ and $\mathcal{H}_{bath}$ respectively. We choose $\{|J_i \rangle_{journal}\}$  to be orthonormal. Then the purified state can be expressed as
\begin{equation} \label{purestate}
|\Psi \rangle = \sum_{i} \sqrt{P(J_i)} \left(\sum_{k,k'} A_{k,k'} |\psi_k , J_i \rangle_{BH}|\gamma_{k'}\rangle_{bath} \right)|J_i\rangle_{journal}. 
\end{equation}
Each $|J_i\rangle_{journal}$ corresponds to a choice of boundary condition for $\text{CFT}_2$ at the physical boundary of the eternal black hole spacetime. \par

We take the CFT$_2$ to be a free theory of  $c_2$ non-compact bosons $X_1,\cdots, X_{c_2}$ with action
\begin{equation}
\label{CFT2S}
S=\sum_{i=1}^{c_2}\frac{1}{2\pi }\int d^2w\partial X_i\bar\partial X_i.
\end{equation}
For this theory the boundary condition $J_i$ corresponds to the boundary value of the boson $X_i$ at the AdS boundary.  We also assume that the boundary conditions are drawn from a Gaussian distribution having standard deviation $1/\delta$ 
\begin{equation}
\label{gaussian}
P\left(J\right)=\frac{\delta}{\sqrt{2\pi}}e^{-\frac{\delta^2}{2}J^2}.
\end{equation}
 It was shown in \cite{Almheiri:2021jwq} that for an eternal black hole the boundary time evolution produces entanglement growth between the black hole and the journal.   This leads to an unbounded logarithmic growth of the journal entropy, producing a unitarity paradox. This information paradox was resolved by introducing an island for the journal which includes the interior of the black hole.

\section{The entanglement wedge of journal}

In this section, we shall determine the entanglement wedge of the journal at late times, which is the main goal of this paper. \par
\label{sec:entwedge of journal}
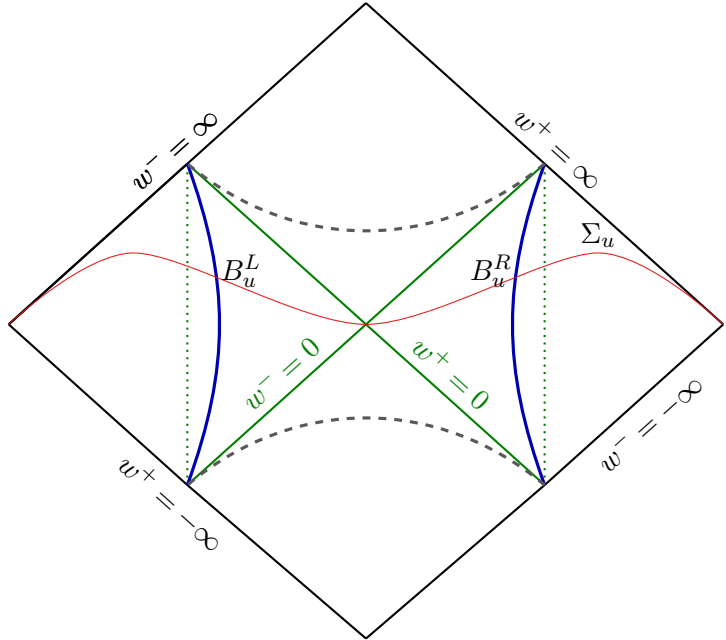
\begin{figure}
\centering
\begin{tikzpicture}[scale=.95]
    \draw [black!50!green,dotted, thick](1 mm, 10 pt) (-1,.25)--(-1,4.75) ;
      \draw [black!50!green,dotted, thick](1 mm, 10 pt) (-6,.25)--(-6,4.75) ;
  \draw [color=black!50!green,thick](1 mm, 10 pt) (-6,4.75)--(-1,.25) node[pos=.7,sloped, above] {$w^+=0$};
 \draw [color=black!50!green,thick](1 mm, 10 pt) (-6,.25)--(-1,4.75)node[pos=.3,sloped, above] {$w^-=0$};
  \draw [thick,](1 mm, 10 pt) (1.5,2.5)--(-1,.25) node[pos=.5,sloped, below] {$w^-=-\infty$} ;
    \draw [thick,](1 mm, 10 pt) (-6,4.75)--(-3.5,7) ;
    \draw [thick,](1 mm, 10 pt) (-1,4.75)--(-3.5,7) ;
      \draw [thick,](1 mm, 10 pt) (-6,.25)--(-3.5,-1.9) ;
    \draw [thick,](1 mm, 10 pt) (-1,.25)--(-3.5,-1.9) ;
 \draw [thick](1 mm, 10 pt) (1.5,2.5)--(-1,4.75)node[pos=1,sloped, above] {$w^+=\infty$} ;
   \draw [thick,](1 mm, 10 pt) (-8.5,2.5)--(-6,.25) node[pos=1,sloped, below] {$w^+=-\infty$} ;
 \draw [thick](1 mm, 10 pt) (-8.5,2.5)--(-6,4.75)node[pos=1,sloped, above] {$w^-=\infty$} ;
  \draw [thick](1 mm, 10 pt) (-8.5,2.5)--(-6,4.75)node[pos=1,sloped, above] {$w^-=\infty$} ;
 \draw [color=black!30!blue, very thick]   (-1,.25)  to[out=110,in=-110] (-1,4.75);
 \draw [color=black!30!blue, very thick]   (-6,.25)  to[in=290,out=-290] (-6,4.75);
  \draw [color=black!30!gray, dashed,very thick]   (-1,4.75)  to[out=-140,in=-40] (-6,4.75) ;
 \draw [color=black!30!gray, dashed, very thick]   (-1,.25)  to[in=40,out=140] (-6,.25);
\draw [red] plot [smooth, very thick] coordinates { (-8.5,2.5) (-6.75,3.5)(-3.5,2.5) (-.25,3.5)(1.5,2.5)};
\draw (-.25,3.75) node {$\Sigma_u$};
\draw (-1.75,3.25) node {$B_{u}^R$};
\draw (-5.25,3.25) node {$B_{u}^L$};
\end{tikzpicture}
\caption{The equal time slice $\Sigma_u$ in the glued spacetime intersects the right AdS boundary at $B_u^R$ and the left AdS boundary at $B_u^L$. } \label{fig:BHbath}
\end{figure}
\subsection{Black hole quantum extremal surfaces}

Consider the setup described in section \ref{sec:setup}. The early time von Neumann entropy of the black hole density matrix can be computed using the replica trick and is given by
\begin{equation}
    \mathbf{S}_{BH}(u) = S^1_{BH}(u) + S^2_{BH}(u),
    \label{eq:initialBHentropy}
\end{equation}
where $S^1_{BH}(u)$ and $S^2_{BH}(u)$ are the CFT$_1$ and CFT$_2$ contributions to the black hole entropy. The CFT$_1$ contribution can be obtained from the twist operator correlation function as follows
\begin{equation}
    {S}^1_{BH}(u) =-\lim_{n\to 1}\frac{1}{1-n} \text{ln} \langle \sigma_1(B_u^L)  \sigma_1(B_u^R)\rangle.
\end{equation}
Here $\sigma_1$ denotes the twist fields in the orbifold version of CFT$_1$ having scaling dimension of $\Delta_n=\frac{c_1}{12}\left(n-\frac{1}{n}\right)$.  The points $B_u^L$ and $B_u^R$ are points on the left and right boundary where the equal time slice $\Sigma_u$ corresponding to the boundary time $u$  intersects the left and right black hole boundaries, see figure \ref{fig:BHbath}. As $u$ increases, it is assumed that the point $B_u^R$ moves along the positive time direction of the right boundary  and the point $B_u^L$ is moving along the negative time direction of the left boundary. The correlation function $\langle \sigma_1(B_u^L)  \sigma_1(B_u^R)\rangle$ is evaluated on a complex plane with the $w$-coordinates described in the previous section with metric $ds^2=\frac{4dw^-dw^+}{\left(1+w^-w^+\right)^2}$. It can be evaluated by  Weyl transforming it into a correlation function on the $w$-plane with flat metric. The resulting correlation function is 
\begin{equation}
\label{twistCFT1corr}
\langle \sigma_1({B_u^L})\rangle  \sigma_1(B_u^R)\rangle=\left(\frac{\left(1+w_{B^L_u}^-w^+_{B^L_u}\right)\left(1+w_{B^R_u}^-w^+_{B^R_u}\right)}{4\left(w_{B^R_u}^+-w^+_{B^L_u}\right)\left(w_{B^R_u}^+-w^+_{B^L_u}\right)}\right)^{\Delta_n}.
\end{equation}
Substituting the $w$-coordinates $(w^+,w^-)$  of the point $B_u^L$ and $B_u^R$ given by
\begin{align}
(w^+_{B_u^L},w^-_{B_u^L})&=\left(-e^{-\frac{2\pi (u-\epsilon)}{\beta}},e^{\frac{2\pi (u+\epsilon)}{\beta}}\right),\nonumber\\
(w^+_{B_u^R},w^-_{B_u^R})&=\left(e^{\frac{2\pi (u-\epsilon)}{\beta}},-e^{\frac{-2\pi (u+\epsilon)}{\beta}}\right)
\end{align}
gives the CFT$_1$ contribution to the black hole entropy as
\begin{equation}
\label{sbhcft1}
{S}^1_{BH}(u) = \frac{c_1}{3}\text{ln}\left(\frac{\beta}{\pi\epsilon}\text{cosh}\left(\frac{2\pi u}{\beta}\right)\right).
\end{equation}
The CFT$_2$ contribution to the black hole entropy is obtained from  the correlation function of the boundary condition changing operators averaged over the distribution $P(\mathbf{J})$ as follows
\begin{equation}
    {S}^2_{BH}(u) =- \lim_{n\to 1}\frac{1}{1-n} \text{ln}\left( \int_{\mathbf{J}^1,\cdots, \mathbf{J}^n}\prod_{i=1, k=1}^{c_2,n}dJ^k_iP(J^{k}_i)\langle \mathcal{O}_{\mathbb{J}}(B_u^L) \mathcal{O}_{\mathbb{J}}(B_u^R)\rangle\right).
  \label{eq:initialBHentropy2}
\end{equation}
The operator $\mathcal{O}_{\mathbb{J}}$ denotes the boundary condition changing operator that changes the boundary conditions of the CFT$_2$ fields. It  changes the boundary conditions of the fields $\{X_1,\cdots,X_{c_2}\}$ from $\mathbf{J}^k=\{J^k_1,\cdots, J^k_{c_2}\}$ to $\mathbf{J}^{k+1}=\{J^{k+1}_1,\cdots, J^{k+1}_{c_2}\}$ as we go from the {$k$-th} sheet to the {$(k+1)$-th} sheet of the replica manifold for $k=1,\cdots,n$. The scaling dimension  of  $\mathcal{O}_{\mathbb{J}}$  is given by
\begin{equation}
\label{scalingdim}
\Delta_{\mathbb{J}}=\sum_{i=1,k}^{c_2,n}\left(\frac{J^{k+1}_i-J^{k}_i}{2\pi}\right)^2.
\end{equation}
  The cut out of AdS$_2$ region is a region in the disk with $w$-coordinates.  The CFT$_2$ correlation function $\langle \mathcal{O}_{\mathbb{J}}(B_u^L) \mathcal{O}_{\mathbb{J}}(B_u^R)\rangle$ is calculated on the $w$-disk and can be obtained with the help of the doubling trick. Finally, the integration over the boundary conditions $\mathbf{J}$ can be performed by using the concept of circularly invariant matrices. For details about circulant matrices see appendix \ref{Circularly invariant matrix}. The detailed integration is described  in appendix \ref{integration}. The final result is given by
  \begin{equation}
  \label{jentropy}
  {S}^2_{BH}(u) \approx \frac{c_2}{2}\text{ln}\left(\frac{u}{\beta\delta^2}\right).
  \end{equation}
Therefore, the black hole entropy at late times has unbounded growth as given below
\begin{equation}
    {S}^1_{BH}(u) \approx \frac{2\pi c_1}{3\beta }u+\frac{c_2}{2}\text{ln}\left(\frac{u}{\beta\delta^2}\right).
\end{equation}
Clearly, such a growth will lead us to information paradox at late times. The resolution of this information paradox requires determining the quantum extremal surface (QES) associated with the black hole. This is done by minimising the generalised entropy of the black hole after removing an interval $A^L_uA^R_u$ from the restriction of the equal time slice $\Sigma_u$ to the AdS$_2$ region. The generalised entropy of the black hole for the interval $B_u^LA_u^L\cup B_u^RA_u^R$ is given by
\begin{equation}
\label{sgenbh}
\mathbf{S}_{BH}^{gen}(u)=\frac{\phi\left(A_u^L\right)+\phi\left(A_u^R\right)}{4G_N}+S_{BH}^{gen,1}(u)+S_{BH}^{gen,2}(u),
\end{equation}
where the first term is the area term. The area term is equal to the sum of  the value of dilaton field given in (\ref{eq:vacuumdilprofw}) at points $A_u^L$ and $A_u^R$.  $S_{BH}^{gen,1}(u)$ and $S_{BH}^{gen,2}(u)$ denote the CFT$_1$ and CFT$_2$ contributions to the generalised black hole entropy. The CFT$_1$ contribution is 
\begin{align}
\label{sgenbh1}
S_{BH}^{gen,1}(u)&=-\lim_{n\to 1}\frac{1}{1-n}\text{ln}\langle \sigma_1\left(B_u^L\right) \sigma_1\left(A_u^L\right) \sigma_1\left(A_u^R\right) \sigma_1\left(B_u^R\right)\rangle\nonumber\\
&\approx -\lim_{n\to 1}\frac{1}{1-n}\text{ln}\langle \sigma_1\left(B_u^L\right) \sigma_1\left(A_u^L\right)\rangle\langle \sigma_1\left(A_u^R\right) \sigma_1\left(B_u^R\right)\rangle \nonumber\\
&\approx \frac{c_1}{6}\text{ln}\left(\left(\frac{\beta}{\pi \epsilon}\right)^2 \frac{\left(e^{-\frac{2\pi u}{\beta}}+w_{A_u^L}^+\right)\left(e^{\frac{2\pi u}{\beta}}-w_{A_u^L}^-\right)}{\left(1+w^-_{A_u^L}w^+_{A_u^L}\right)}\right) \left( \frac{\left(e^{-\frac{2\pi u}{\beta}}+w_{A_u^R}^-\right)\left(e^{\frac{2\pi u}{\beta}}-w_{A_u^R}^+\right)}{\left(1+w^-_{A_u^R}w^+_{A_u^R}\right)}\right).
\end{align}
In the second step we made the approximation by assuming that the points $A_u^L$ and $A_u^R$ are well separated. 

The CFT$_2$ contribution is given by
\begin{align}
\label{sgenbh2}
S_{BH}^{gen,2}(u)&=-\lim_{n\to 1}\frac{1}{1-n}\text{ln}\int_{\mathbf{J}^1,\cdots, \mathbf{J}^n}\prod_{i=1, k=1}^{c_2,n}dJ^k_iP(J^{k}_i) \langle \mathcal{O}_{\mathbb{J}}\left(B_u^L\right) \sigma_2\left(A_u^L\right) \sigma_2\left(A_u^R\right) \mathcal{O}_{\mathbb{J}}\left(B_u^R\right)\rangle\nonumber\\
&\approx-\lim_{n\to 1}\frac{1}{1-n}\text{ln}\int_{\mathbf{J}^1,\cdots, \mathbf{J}^n}\prod_{i=1, k=1}^{c_2,n}dJ^k_iP(J^{k}_i)\langle \mathcal{O}_{\mathbb{J}}\left(B_u^L\right) \sigma_2\left(A_u^L\right)\rangle\langle \sigma_2\left(A_u^R\right) \mathcal{O}_{\mathbb{J}}\left(B_u^R\right)\rangle,
\end{align}
where $\sigma_2$ denotes the CFT$_2$ twist operators. The correlators $\langle \mathcal{O}_{\mathbb{J}}\left(B_u^L\right) \sigma_2\left(A_u^L\right)\rangle$ and $\langle \sigma_2\left(A_u^R\right) \mathcal{O}_{\mathbb{J}}\left(B_u^R\right)\rangle$ are evaluated on the $w$-plane where CFT$_2$ is defined. Since in the Euclidean version this region is a cutout of disk, these correlators can be calculated by using the doubling trick. The correlator $\left\langle \mathcal{O}_{\mathbb{J}}\left(w_{B_u^L}^+,w_{B_u^L}^-\right) \sigma_2\left(w_{A_u^L}^+,w_{A_u^L}^-\right)\right\rangle$ evaluated on the Euclidean AdS$_2$ is given by
\begin{equation}
\label{osigma}
\left\langle \mathcal{O}_{\mathbb{J}}\left(w_{B_u^L}^+,w_{B_u^L}^-\right) \sigma_2\left(w_{A_u^L}^+,w_{A_u^L}^-\right)\right\rangle=G_n\left(\mathbb{J} \right)\left(\left(\frac{\pi \epsilon}{\beta}\right)\frac{\left(1+w_{A_u^L}^-w_{A_u^L}^+\right)}{\left(1-e^{-\frac{2\pi u}{\beta}}w_{A_u^L}^-\right)\left(1+e^{\frac{2\pi u}{\beta}}w_{A_u^L}^+\right)}\right)^{\Delta_{\mathbb{J}}}.
\end{equation}
The coefficient $G_n\left(\mathbb{J} \right)$ is related to the $n$-point function of boundary condition changing operators kept on a disk
\begin{equation}
G_n\left(\mathbb{J} \right)=\prod_{k\neq l,i=1}^{c_2}\left|e^{\frac{2\pi {{i}}(k-1)}{n}}-e^{\frac{2\pi {i}(l-1)}{n}}\right|^{\mu^i_{kl}},
\label{corrboson1}
\end{equation}
where $\mu^i_{kl}=\frac{\left({J}_i^{{k+1}}-{J}_i^{k}\right)\left({J}_i^{{l+1}}-{J}_i^{l}\right)}{2\pi^2}$. After performing the averaging over the Gaussian distribution again by using the integration method based on circulant matrix we obtain the generalised bath entropy as follows  
\begin{align}
\label{sgenbh3}
S_{BH}^{gen,2}(u)\approx \frac{c_2}{2}\text{ln}\left(\text{ln}\left(\left(\frac{\beta}{\pi \epsilon}\right)^2\frac{\left(1-e^{-\frac{2\pi u}{\beta}}w_{A_u^L}^-\right)\left(1+e^{\frac{2\pi u}{\beta}}w_{A_u^L}^+\right)\left(1+e^{\frac{2\pi u}{\beta}}w_{A_u^R}^-\right)\left(1-e^{-\frac{2\pi u}{\beta}}w_{A_u^R}^+\right)}{\left(1+w_{A_u^L}^-w_{A_u^L}^+\right)\left(1+w_{A_u^R}^-w_{A_u^R}^+\right)}\right)\right).
\end{align}
For more details about this computation, see appendix \ref{integration}. Extremising  the  generalised bath entropy  with respect to  $w_{A_u^L}^+$ and  $w_{A_u^L}^-$ gives the following QES equations
\begin{align}
-\frac{\pi \phi_r}{G_N\beta}\frac{w_{A_u^L}^-}{\left(1+w^+_{A_u^L}w^-_{A_u^L}\right)^2}+\left(\frac{c_1}{6}+\frac{c_2}{2~\text{ln}\left(\frac{\beta}{\pi\epsilon}\right)}\right)\left(\frac{1}{e^{-\frac{2\pi u}{\beta}}+w^+_{A_u^L}}-\frac{w_{A_u^L}^-}{\left(1+w^+_{A_u^L}w^-_{A_u^L}\right)}\right)&=0\nonumber\\
-\frac{\pi \phi_r}{G_N\beta}\frac{w_{A_u^L}^+}{\left(1+w^+_{A_u^L}w^-_{A_u^L}\right)^2}-\left(\frac{c_1}{6}+\frac{c_2}{2~\text{ln}\left(\frac{\beta}{\pi\epsilon}\right)}\right)\left(\frac{1}{e^{\frac{2\pi u}{\beta}}-w^-_{A_u^L}}+\frac{w_{A_u^L}^+}{\left(1+w^+_{A_u^L}w^-_{A_u^L}\right)}\right)&=0.
\end{align} 
There exists  a solution for this coupled equations at late times near the left future horizon of the black hole where $w^+_{A_u^L}w^-_{A_u^L}\approx 0$ . The solution is given by 
\begin{align}
w_{A_u^L}^{\pm}=\mp \frac{G_N\beta}{6\pi \phi_r}\left(c_1+\frac{3c_2}{\text{ln}\left(\frac{\beta}{\pi\epsilon}\right)}\right)e^{\mp\frac{2\pi u}{\beta}}.
\end{align}
By repeating the same analysis we can find the QES in the right side of the black hole. It is given by 
\begin{align}
w_{A_u^R}^{\pm}=\pm \frac{G_N\beta}{6\pi \phi_r}\left(c_1+\frac{3c_2}{\text{ln}\left(\frac{\beta}{\pi\epsilon}\right)}\right)e^{\pm\frac{2\pi u}{\beta}}.
\end{align}
Substituting the QES solutions back to the generalised black hole entropy expression shows that at late time the black hole entropy becomes a constant equal to twice the area of black hole horizon. Therefore, this QES after Page time $u_{Page}$,  tame the non-unitary growth of the black hole entropy.

\subsection{Entanglement wedge of bath and the extended strong subadditivity}

The QES computation in the previous subsection suggests that after Page time $u_{Page}\approx \frac{3S^0_{BH}\beta}{\pi c_1}$, where $S^0_{BH}$ is area of the bifurcation horizon of the black hole,  the combined system of the bath and the journal possesses a non-trivial entanglement wedge that contains the interior of the black hole. The entanglement wedge of the journal at late time must belong to the entanglement wedge of the combined system. Hence, we should search for a journal island satisfying the constraints of the extended strong subadditivity  \cite{carlen2012bounds}  inside the interval bounded by the black hole quantum extremal surfaces. The eSSA is an inequality satisfied by the von Neumann entropies of three subsystems of  a larger quantum system which we explain below. \par 

Consider a quantum system having  Hilbert space $\mathcal{H}$ formed by taking the tensor product of the Hilbert spaces of three subsystems $\mathcal{H}_1,\mathcal{H}_2$ and $\mathcal{H}_3$, i.e. $\mathcal{H}=\mathcal{H}_1\otimes\mathcal{H}_2\otimes\mathcal{H}_3$. We denote the state of the larger quantum system by $\rho^{123}$, the state of the combined system having Hilbert space $\mathcal{H}^{ij}=\mathcal{H}^i\otimes \mathcal{H}^j$  by $\rho^{ij}$, and the state of the $i^{th}$ subsystem having Hilbert space $\mathcal{H}^i$  by $\rho^i$. Then the eSSA inequality states that 
\begin{equation}
\label{improvedSSA}
S\left(\rho^{12}\right)+S\left(\rho^{23}\right)-S\left(\rho^{123}\right)-S\left(\rho^{2}\right)\geq 2~max\left\{S\left(\rho^{1}\right)-S\left(\rho^{13}\right),S\left(\rho^{3}\right)-S\left(\rho^{13}\right),0 \right\};
\end{equation}
for  the usual strong subadditivity inequality the right hand side is simply zero. \par

Now we make use of the above inequality as follows. Take subsystem 1 to be the journal, subsystem 2 to be the black hole and the subsystem 3 to be the bath. Then the eSSA satisfied by the  entropies of the subsystems after Page time reads as follows 
\begin{equation}
\label{improvedSSABHJB}
S_{bath}(u)+S_{journal}(u)-2S^0_{BH}\geq 2~max\left\{S_{journal}(u)-2S_{BH}^0, S_{bath}(u)-2S_{BH}^0,0\right\}.
\end{equation}
Using the replica method the entropies of bath and journal can be calculated. The bath entropy at large times is given by
\begin{align}
    {S}_{bath}(u)&=-\lim_{n\to 1}\frac{1}{1-n} \text{ln} \langle \sigma_1(B_u^L)  \sigma_1(B_u^R)\rangle \approx\frac{2\pi c_1}{3\beta}u,
\end{align}
and the journal entropy is given by
\begin{equation}
    {S}_{journal}(u) =- \lim_{n\to 1}\frac{1}{1-n} \text{ln}\left( \int_{\mathbf{J}^1,\cdots, \mathbf{J}^n}\prod_{i=1, k=1}^{c_2,n}dJ^k_iP(J^{k}_i)\langle \mathcal{O}_{\mathbb{J}}(B_u^L) \mathcal{O}_{\mathbb{J}}(B_u^R)\rangle\right)\approx \frac{c_2}{2}\text{ln}\left(\frac{u}{\beta\delta^2}\right).
  \label{jouralE}
\end{equation}
After Page time, since the bath entropy is significantly greater than the entropy of the journal, the eSSA takes the following form
\begin{equation}
\label{improvedSSABHJBRPage}
S_{bath}(u)-S_{journal}(u)\leq 2S^0_{BH}.
\end{equation}
It is clear from these expressions that the entropy of the bath and the journal violates the eSSA (\ref{improvedSSABHJBRPage}) after time $u=u_B>u_{Page}$, where $u_B$ is the time at which the difference in  $S_{bath}(u)$ and $S_{journal}(u)$ becomes $2S^0_{BH}$. The root cause of this violation is the linear growth of entanglement entropy of the bath while the journal only has logarithmic growth of entanglement entropy. Therefore, this violation of eSSA can be described as the bath information paradox. \par

An island for bath that is inside the interval enclosed by the quantum extremal surfaces of black hole might resolve this paradox. With this hope, let us search for a bath island by minimising the  generalised entropy of bath associated with an arbitrary interval at $C_u^LC_u^R$ with respect to the points $C_u^L$ and $C_u^R$.  The generalised entropy of the bath is given by
\begin{equation}
\label{bathgentropy}
S_{bath}^{gen}=\frac{\phi\left(C_u^L\right)+\phi\left(C_u^R\right)}{4G_N}+S_{bath}^{gen,1}(u)+S_{bath}^{gen,2}(u),
\end{equation}
where $S_{bath}^{gen,1}(u)$ is the CFT$_1$ contribution to the bath generalised entropy
\begin{align}
\label{sgenbath1}
S_{bath}^{gen,1}(u)&=-\lim_{n\to 1}\frac{1}{1-n} \text{ln} \langle\sigma_1(B_u^L)\sigma_1(C_u^L)\sigma_1(C_u^R)\sigma_1(B_u^R)\rangle\nonumber\\
&\approx -\lim_{n\to 1}\frac{1}{1-n} \text{ln} \langle\sigma_1(B_u^L)\sigma_1(C_u^L)\rangle\langle\sigma_1(C_u^R)\sigma_1(B_u^R)\rangle\nonumber\\
&\approx \frac{c_1}{6}\text{ln}\left(\left(\frac{\beta}{\pi \epsilon}\right)^2 \frac{\left(e^{-\frac{2\pi u}{\beta}}+w_{C_u^L}^+\right)\left(e^{\frac{2\pi u}{\beta}}-w_{C_u^L}^-\right)}{\left(1+w^-_{C_u^L}w^+_{C_u^L}\right)}\right) \left( \frac{\left(e^{-\frac{2\pi u}{\beta}}+w_{C_u^R}^-\right)\left(e^{\frac{2\pi u}{\beta}}-w_{C_u^R}^+\right)}{\left(1+w^-_{C_u^R}w^+_{C_u^R}\right)}\right),
\end{align}
and  $S_{bath}^{gen,2}(u)$ is the CFT$_2$ contribution to the bath generalised entropy
\begin{align}
\label{sgenbath2}
S_{bath}^{gen,2}(u)&=-\lim_{n\to 1}\frac{1}{1-n} \text{ln} \int_{\mathbf{J}^1,\cdots, \mathbf{J}^n}\prod_{i=1, k=1}^{c_2,n}dJ^k_iP(J^{k}_i)\langle\sigma_2(C_u^L)\sigma_2(C_u^R)\rangle\nonumber\\
&\approx -\lim_{n\to 1}\frac{1}{1-n} \text{ln} \int_{\mathbf{J}^1,\cdots, \mathbf{J}^n}\prod_{i=1, k=1}^{c_2,n}dJ^k_iP(J^{k}_i) \langle\sigma_2(C_u^L)\rangle\langle\sigma_2(C_u^R)\rangle\approx 0.
\end{align}
Here we used the fact that any one point function of a primary field in AdS$_2$ disk is identity.  The solution for the associated coupled QES equations at late time is given by
\begin{align}
w_{C_u^L}^{\pm}=\mp \frac{G_N\beta}{6\pi \phi_r}c_1e^{\mp\frac{2\pi u}{\beta}}, \qquad w_{C_u^R}^{\pm}=\pm \frac{G_N\beta}{6\pi \phi_r}c_1e^{\pm\frac{2\pi u}{\beta}}.
\end{align}
Substituting the QES solutions back into the expression for the generalised bath entropy (\ref{bathgentropy}) gives a constant equal to twice the area of black hole horizon, i.e. 
\begin{equation}
S^{gen}_{bath}(u)=2S^0_{BH}, \qquad u>u_B.
\end{equation}

We must check whether the island with boundaries that matches with the above quantum extremal surfaces resolves the bath information paradox that appeared soon after black hole Page time $u_{Page}$. The eSSA inequality  (\ref{improvedSSABHJBRPage}) is satisfied after the  black hole Page time if we replace with $S_{bath}(u)$  with $S^{gen}_{bath}(u)$, as long as $S_{journal}(u) \leq S^{gen}_{bath}(u) \sim  2 S^0_{BH}$. Until $u=u_I$ at which  $S_{journal}(u_I)= 2S^0_{BH}$, the eSSA reduces to the demand that $S_{journal}(u)\geq 0$.  Thus, the inclusion of the bath island enables all the three subsystems to satisfy the eSSA inequality  (\ref{improvedSSABHJBRPage}) at least till $u=u_I$. 

\subsection{Transfer of the ownership of the black hole interior from bath to journal}

After time $u=u_I$, while the bath owns a non-trivial island  that contains the black hole interior,  the eSSA (\ref{improvedSSABHJB})  is given by 
\begin{equation}
\label{improvedSSABHJBRPage1}
S_{journal}(u)-S^{gen}_{bath}(u)\leq 2S^0_{BH}, \qquad u>u_I.
\end{equation}
It is straightforward to see that this inequality will be violated  after time $u=u_J$, at which $S_{journal}(u_J)= 4S^0_{BH}$. At late times, for $u>u_J$, this leads to another unitarity violation or  information paradox. In order to resolve this paradox, we shall search for an island for the journal by minimising the  generalised entropy of journal associated with an arbitrary interval at $D_u^LD_u^R$. The generalised entropy of the journal associated with this interval is given by
\begin{equation}
\label{journalgentropy}
S_{bath}^{gen}=\frac{\phi\left(D_u^L\right)+\phi\left(D_u^R\right)}{4G_N}+S_{journal}^{gen,1}(u)+S_{journal}^{gen,2}(u).
\end{equation}
Here $S_{journal}^{gen,1}(u)$ is the CFT$_1$ contribution to the bath generalised entropy 
\begin{align}
\label{sgenjournal1}
S_{journal}^{gen,1}(u)&=-\lim_{n\to 1}\frac{1}{1-n} \text{ln} \langle\sigma_1(D_u^L)\sigma_1(D_u^R)\rangle \approx \frac{c_1}{6}\text{ln}\left(\frac{\left(w_{D_u^R}^+-w_{D_u^L}^+\right)\left(w_{D_u^R}^--w_{D_u^L}^-\right)}{\left(1+w^-_{D_u^L}w^+_{D_u^L}\right)\left(1+w^-_{D_u^R}w^+_{D_u^R}\right)}\right)
\end{align}
and  $S_{journal}^{gen,2}(u)$ is the CFT$_2$ contribution to the bath generalised entropy. 
\begin{align}
\label{sgenjournal2}
S_{journal}^{gen,2}(u)&=-\lim_{n\to 1}\frac{1}{1-n} \text{ln} \int_{\mathbf{J}^1,\cdots \mathbf{J}^n}\prod_{i=1, k=1}^{c_2,n}dJ^k_iP(J^{k}_i)\langle\mathcal{O}_{\mathbb{J}}(B_u^L)\sigma_2(D_u^L)\sigma_2(D_u^R)\mathcal{O}_{\mathbb{J}}(B_u^R)\rangle\nonumber\\
&\approx -\lim_{n\to 1}\frac{1}{1-n} \text{ln} \int_{\mathbf{J}^1,\cdots \mathbf{J}^n}\prod_{i=1, k=1}^{c_2,n}dJ^k_iP(J^{k}_i) \langle\mathcal{O}_{\mathbb{J}}(B_u^L)\sigma_2(D_u^L)\rangle\langle\sigma_2(D_u^R)\mathcal{O}_{\mathbb{J}}(B_u^R)\rangle\nonumber\\
&\approx \frac{c_2}{2}\text{ln}\left( \text{ln}\left(\left(\frac{\beta}{\pi \epsilon}\right)^2\frac{\left(1-e^{-\frac{2\pi u}{\beta}}w_{D_u^L}^-\right)\left(1+e^{\frac{2\pi u}{\beta}}w_{D_u^L}^+\right)\left(1+e^{\frac{2\pi u}{\beta}}w_{D_u^R}^-\right)\left(1-e^{-\frac{2\pi u}{\beta}}w_{D_u^R}^+\right)}{\left(1+w_{D_u^L}^-w_{D_u^L}^+\right)\left(1+w_{D_u^R}^-w_{D_u^R}^+\right)}\right)\right).
\end{align}

The generalised entropy minimisation with respect to  $w_{D_u^L}^+$ and  $w_{D_u^L}^-$ gives the following equations
\begin{align}
\label{QESjournal}
-\frac{\pi \phi_r}{G_N\beta}\frac{w_{D_u^L}^-}{\left(1+w^+_{D_u^L}w^-_{D_u^L}\right)^2}+\frac{c_2}{2~\text{ln}\left(\frac{\beta}{2\pi\epsilon}\right)}\left(\frac{1}{e^{-\frac{\pi u}{\beta}}+w^+_{D_u^L}}-\frac{w_{D_u^L}^-}{\left(1+w^+_{D_u^L}w^-_{D_u^L}\right)}\right)&=0\nonumber\\
-\frac{\pi \phi_r}{G_N\beta}\frac{w_{D_u^L}^+}{\left(1+w^+_{D_u^L}w^-_{D_u^L}\right)^2}-\frac{c_2}{2~\text{ln}\left(\frac{\beta}{2\pi\epsilon}\right)}\left(\frac{1}{e^{\frac{\pi u}{\beta}}-w^-_{D_u^L}}+\frac{w_{D_u^L}^+}{\left(1+w^+_{D_u^L}w^-_{D_u^L}\right)}\right)&=0.
\end{align} 
The solution near horizon at late time is given by
\begin{align}
w_{D_u^L}^{\pm}&=\mp \frac{G_N\beta c_2}{2\pi \phi_r \text{ln}\left(\frac{\beta}{2\pi\epsilon}\right)}e^{\mp\frac{2\pi u}{\beta}}\\\nonumber
w_{D_u^R}^{\pm}&=\pm \frac{G_N\beta c_2}{2\pi \phi_r \text{ln}\left(\frac{\beta}{2\pi\epsilon}\right)}e^{\pm\frac{2\pi u}{\beta}}.
\end{align}
Substituting this back to the expression (\ref{journalgentropy}) gives the generalised entropy of the subsystem journal as follows
\begin{align}
\label{journalgentropy1}
S_{journal}^{gen}(u)&=2S^0_{BH}+\frac{c_1}{3}\text{ln}\left[\frac{\beta}{\pi\epsilon}\text{cosh}\left(\frac{2\pi u}{\beta}\right)\right] \nonumber\\
&=2S^0_{BH}+S_{bath}(u).
\end{align}
Let us allow the journal to own the island  that contains the black hole interior instead of the bath after time $u=u_J$. Then the eSSA relation for  $u>u_J$  is given by 
\begin{equation}
\label{improvedSSABHJBRPage2}
S_{journal}^{gen}(u)-S_{bath}(u)\leq 2S^0_{BH}, \qquad u>u_{J}.
\end{equation}
Using  (\ref{journalgentropy1}) we can verify that this eSSA relation gets saturated after time $u=u_J$. Therefore,  transferring the ownership of the black hole interior after time $u=u_J$ from the bath to the journal restores unitarity.  Consequently, the reconstruction of the black hole interior from radiation at late times requires complete knowledge of the theory.

\section{Conclusion}
\label{conclusion}
We analysed the theory dependence of the interior reconstruction of an AdS$_2$ eternal black hole in equilibrium with a finite temperature bath by introducing a CFT with matter fields having random reflecting boundary conditions along the AdS$_2$ boundaries. By using the island formula and the extended strong subadditivity due to Carlen and Lieb, we have shown that at late times the reference Hilbert space that encodes the information about the random boundary conditions owns an entanglement wedge that contains the black hole interior including the region near singularity. This implies that  the reconstruction of the region near singularity of a black hole from radiation requires exquisite knowledge of the theory. \par
One interesting point to note is that the combined system of black hole, bath and the journal before Page time was in a state that saturated the  extended strong subadditivity  \cite{Hayden:2004sss}. Interestingly the state of the combined system that satisfies all the unitarity requirements at late time saturates the extended strong subadditivity. It would be interesting to study the significance of this observation. As already pointed out in \cite{Almheiri:2021jwq}, extending this analysis to an evaporating black hole can teach us about the theory dependence of the black hole interior reconstruction from  Hawking radiation.

\section*{Acknowledgements}

We thank  Sujay Ashok, Yiming Chen, Sibasish Ghosh, Rajesh Gopakumar, Thomas Hartman, Alok Laddha, Ayan Mukopadhyay, Onkar Parrikar, Loganayanam R,  Suvrat Raju,  Mukund Rangamani, Anupam Sarkar, and Ashoke Sen for the valuable discussions and suggestions.


\appendix

\section{Circulant matrices} 
\label{Circularly invariant matrix}
 A  circulant matrix $A$ is a square matrix with the property that  all its rows are made up of the same elements and each row is undergoing a cyclic shift of one element to the right relative to the preceding row
 \begin{equation}
A= \begin{bmatrix}
a_0 & a_1 & a_2& &\cdots & a_{n-1}\\
a_{n-1} & a_0 & a_1 & &\cdots& \vdots \\
 & a_{n-1} & a_0 & a_1& \ddots &\\ 
\vdots& &\ddots& \ddots & \ddots & a_2\\
 &&&&a_1\\
 a_1 & \cdots & &  & a_{n-1} &a_0
\end{bmatrix}.
\label{circularlyinvm}
 \end{equation}

 An interesting feature of a circularly invariant matrix is that for arbitrary values of $n$ its eigenvalues can be obtained \cite{davis1979circulant} and are given by
\begin{equation}
\lambda_p=\sum_{k=1}^{n-1}a_ke^{-\frac{2\pi i pk}{n}} \qquad p=0,\cdots,n-1.
\label{Ceigenv}
\end{equation}
Therefore\VH{,} the determinant of the circulant matrix $A$ is given by
\begin{equation}
det(A)=\prod_{r=0}^{n-1}\left(\sum_{k=1}^{n-1}a_ke^{-\frac{2\pi i rk}{n}}\right).
\label{Ceigenv1}
\end{equation}

 \section{Averaging over the distribution}
\label{integration}
In this appendix, we shall perform the averaging over a Gaussian distribution  of the correlation function 
$$\int_{\mathbf{J}^1,\cdots \mathbf{J}^n}\prod_{i=1, k=1}^{c_2,n}dJ^k_iP(J^{k}_i)\langle \mathcal{O}_{\mathbb{J}}\left(B_u^L\right) \sigma_2\left(A_u^L\right)\rangle\langle \sigma_2\left(A_u^R\right) \mathcal{O}_{\mathbb{J}}\left(B_u^R\right)\rangle.$$
The Gaussian distribution  for the random boundary conditions is given by
$$P\left(J_i^k\right)=\frac{\delta}{\sqrt{2\pi}}e^{-\frac{\delta^2}{2}(J_i^k)^2}.$$
After substituting the expressions for the correlation functions, we get that
\begin{align}
\label{sgenbh4}
&\int_{\mathbf{J}^1,\cdots \mathbf{J}^n}\prod_{i=1, k=1}^{c_2,n}dJ^k_iP(J^{k}_i)\langle \mathcal{O}_{\mathbb{J}}\left(B_u^L\right) \sigma_2\left(A_u^L\right)\rangle\langle \sigma_2\left(A_u^R\right) \mathcal{O}_{\mathbb{J}}\left(B_u^R\right)\rangle\nonumber\\
&=\frac{m^{nc_2}}{(2\pi)^{nc_2}}\prod_{i=1}^{c_2}\left(\int_{{J}_i^1,\cdots {J}_i^n}dJ^1_i\cdots dJ^n_ie^{-\sum_{p,q}A_{pq}J_p^kJ_q^k}\right).
\end{align}
The elements $A_{pq}$ of  the $n\times n$  square matrix $A$ can be expressed as 
 \begin{equation}
A_{pq}=\left(\frac{m^2}{2}-2Q\right)\delta_{p,q}-2e_{pq}\left(1-\delta_{p,q}\right)+e_{p(q-1)}\left(1-\delta_{p,(q-1)}\right)+e_{p(q+1)}\left(1-\delta_{p,(q+1)}\right)+2Q\delta_{p,(q-1)},
 \label{Aascircmatrix}
 \end{equation}
 where $e_{pq}$ is given by  $$e_{pq}=\frac{1}{2\pi^2}\text{ln}\left(4~\sin^2\left(\frac{\pi (p-q)}{n}\right)\right),$$
and $Q$ is given by  $$Q=\frac{1}{4\pi^2}\text{ln}~\left(\left(\frac{\pi \epsilon}{\beta}\right)^2\frac{\left(1+w_{A_u^L}^-w_{A_u^L}^+\right)\left(1+w_{A_u^R}^-w_{A_u^R}^+\right)}{\left(1-e^{-\frac{2\pi u}{\beta}}w_{A_u^L}^-\right)\left(1+e^{\frac{2\pi u}{\beta}}w_{A_u^L}^+\right)\left(1-e^{-\frac{2\pi u}{\beta}}w_{A_uR}^-\right)\left(1+e^{\frac{2\pi u}{\beta}}w_{A_u^R}^+\right)}\right).$$
The matrix elements $A_{pq}$ have the following  shift symmetry 
\begin{equation}
A_{pq}=A_{(\{p+m\})(\{q+m\})}.
\end{equation}
 Here, the curly bracket $\{\}$ in the subscript indicates that $\{p+m\}$ is $p+m$ for $p+m\leq n$  and $n-p-m$ otherwise. This assures that $A$ is a circulant matrix. Therefore, the determinant of the matrix $A$ is given by 
 \begin{equation}
 det(A)=\prod_{r=0}^{n-1}\left(\sum_{k=1}^{n-1}A_{1k}e^{-\frac{2\pi i rk}{n}}\right).
 \label{detA}
 \end{equation}
  Using Gauss's digamma theorem  it is possible to verify that for $\text{ln}(\text{det}(A))$, the limits $n\to 1$ and $Q\to \infty$ commutes.  Gauss's digamma theorem  is the following identity
  \begin{equation}
  \sum_{m=1}^{n-1}e^{-\frac{2\pi i mp}{n}} \ln\left(4~\sin^2\left(\frac{m\pi}{n}\right) \right)=2~\ln~n +2\gamma+2\psi\left(\frac{p}{n} \right)+\pi~\cot\left( \frac{\pi p}{n}\right)  \qquad p=1,\cdots, n-1.
  \end{equation}
  where $\psi$ is the digamma function $\psi(x)=\frac{d}{dx} \ln~ \Gamma(x)$ and $\gamma$ is the Euler's constant. Since at late times $Q$ is very large  we can approximate the logarithm of the  determinant of $A$ at late times as

  \begin{equation}
  \begin{split}
  \ln\left(det\left(A\right)\right) &\sim n~\ln\left( \frac{\delta^{2}}{2}\right)-(n-1)Q+\frac{4\left((n-1)\gamma+n~\ln~n\right)}{Q\pi^2\delta^2}.
 \end{split}
 \label{lndetapprox}
 \end{equation} 
  The third term is the leading contribution from $G_n\left(\mathbb{J}\right)$. This implies that for large values of $Q$, contribution from $G_n\left(\mathbb{J}\right)$ to the integral is negligible and hence can be safely ignored while we perform the integration that does the averaging over the distribution.


\bibliographystyle{JHEP}
\bibliography{reference}
\end{document}